\documentclass[3p,times,11pt]{elsarticle}
\usepackage{setspace}
\usepackage{lipsum}
\usepackage{mdsymbol}
\usepackage{amsthm}
\makeatletter
\def\ps@pprintTitle{%
	\let\@oddhead\@empty
	\let\@evenhead\@empty
	\def\@oddfoot{}%
	\let\@evenfoot\@oddfoot}
\makeatother
\usepackage{lineno,hyperref,booktabs}
\modulolinenumbers[5]

\journal{arXiv preprint}

\begin{document}

\begin{frontmatter}

\title{Fast Cauchy-Rician Modelling of SAR Images \\
with Method of Algebraic Moments Estimator}

\author{Mutong LI}

\author{Ercan Engin KURUOĞLU}

\address{Tsinghua-Berkeley Shenzhen Institute, Tsinghua University}

\begin{abstract}
SAR technology has been intensively implemented for geo-sensing and mapping purposes due to its advantages of high azimuthal resolution and weather-independent operation compared to other remote sensing technologies. Modelling SAR image data consequently becomes a prominent topic of interest, especially for data populations with impulsive signal features, which are common  in SAR images of urban areas. A recently proposed model named Cauchy-Rician has manifested great potential in modelling extremely heterogeneous SAR images, yet the work only provided a MCMC-based parameter estimator that demands considerable computational power. In this work, a novel analytical parameter estimation method based on algebraic moments is proposed to provide stable and accurate estimation of the parameters of the Cauchy-Rician model with significant improvement on computation speed. 

\end{abstract}


\end{frontmatter}

\section{Introduction}
\label{Section: Introduction}
Synthetic aperture radar (SAR) is a powerful radar technology commonly adopted for airborne and spaceborne geo-sensing and mapping applications due to its significant advantages of high azimuthal resolution and weather-independent operation \cite{moreira2013tutorial}. However, the coherent imaging nature of SAR has risen difficulties in the analysis of the radar image, demanding competent models to properly characterize features of image data population produced by various terrains. Amongst these terrains stand out a particularly troublesome case, namely the signal intensities of urban areas which usually assumes an impulsive, heavy-tailed population. This phenomenon is generally attributed to urban areas’ richness in cars and steel-plate rooftops that serves as perfect reflectors to radio-frequency electromagnetic waves, manifesting themselves as extremely bright pixels in the grand picture.

Intensive efforts have been made to develop statistical models for characterizing and analysing SAR image models. The most classical model is the Rayleigh distribution, which models the amplitude of a complex signal by assuming both the real and imaginary parts to be jointly zero-mean Gaussian. The Rayleigh model has succeeded in modelling homogeneous scenes, since Gaussian distribution, as suggested by the Central Limit Theorem, is the limit distribution of a collection of infinitesimal reflectors with finite variance. 
The model was later generalized to the Rician model with the introduction of an additional location parameter to accommodate impulsive radar scenes with a dominant reflector \cite{rice1944mathematical}. However, for urban SAR images of extreme heterogeneity, the scene generally includes not only one but many a dominant reflectors, rendering the Rician model less competent.
This observation lead some researchers to look for alternative impulsive distributions \cite{chen2016} to replace Gaussian distribution in the development of the Rician model. In particular, \cite{karakucs2021generalized} proposed a generalisation of the Rician distribution based on Generalised Gaussian real and complex components, in parallel to extension of Rayleigh distribution in \cite{moser2006}, which showed success in modelling SAR images of the ocean scenes with only modest performance for urban areas due to the mild impulsiveness of the generalised Gaussian distribution. In order to accommodate the heterogeneity observed in urban area data, models based on Cauchy distribution which is itself extremely impulsive has been intensively studied. Despite being theorized in the early 1960s \cite{ferguson1962representation}, the idea of modelling extremely heterogeneous complex (radar) signals with a circular-bivariate Cauchy distribution was proposed only recently \cite{karakucs2022cauchy}, where a model named “Cauchy-Rician” was established to model the amplitude of complex SAR image signals from urban areas. In accordance with the model, a parameter estimator was developed based on Markov Chain Monte Carlo (MCMC). Despite its effectiveness, the MCMC parameter estimator being a sampling based learning method requires a significant amount of computational power, and is consequently slow for real-time data processing. 

In this paper, a novel parameter estimator based on a generalisation of the  method of moments is proposed to resolve the matter. The method utilizes expectations of algebraic functions to produce analytical equations based on which the parameters are solved, since conventional method of moments based on expectations of power functions fail to yield compact results. Advantages of this method of algebraic moments, including fast speed, simplicity, and stability will be revealed in the following parts of the paper. In the second section, we derive an analytically simpler expression for the Cauchy-Rician model. Based on this new expression we propose our parameters estimator based on method of algebraic moments. In Section 3, we report our experimental study on synthetic and real data, respectively. Section 4 provides conclusions and a discussion.

\section{Cauchy-Rician Distribution}
\label{Section: C-R Derivation}

\subsection{Derivation of the Cauchy-Rician model}
The derivation starts with the circular-bivariate Cauchy distribution defined as in \cite{ferguson1962representation}, the basic idea lies in assuming both the real and imaginary parts of a complex image signal to be Cauchy distributed. The characteristic function of a circular-bivariate Cauchy distribution is 
\begin{equation}
    \varphi\left( {\omega_{1},\omega_{2}} \right) = \exp\left\lbrack j\left( \delta_{1}\omega_{1} + \delta_{2}\omega_{2} \right) - \gamma\sqrt{\omega_{1}^{2} + \omega_{2}^{2}} \right\rbrack
\end{equation}
$\gamma$ is the scale parameter, $\delta_{1}$ and $\delta_{2}$ are the location parameters of the real and imaginary parts of a complex signal. By performing the inverse Fourier transform of the above characteristic function one acquires the pdf of circular-bivariate Cauchy.
\begin{equation}
    {pdf}_{X_{re},~X_{im}}\left( {x_{re},~x_{im}} \right) = \frac{1}{4\pi^{2}}{\iint\limits_{\omega_{1},~\omega_{2}}^{~}{{\exp\left\lbrack {j\left( {\delta_{1}\omega_{1} + \delta_{2}\omega_{2}} \right) - \gamma\left| \boldsymbol{\omega} \right|^{\alpha}} \right\rbrack}~{\exp\left\lbrack {j\left( {\omega_{1}x_{re} + \omega_{2}x_{im}} \right)} \right\rbrack}~d\omega_{1}d\omega_{2}}}
\end{equation}
By assuming $\omega=\sqrt{\omega_1^2+\omega_2^2}$ and $\theta_\omega=\arctan{\frac{\omega_2}{\omega_1}}$, together with $x=\sqrt{x_{re}^2+x_{im}^2}$ and $\theta_x=\arctan{\frac{x_{im}}{x_{re}}}$, the above integral can be converted into the polar coordinates.
\begin{equation}
\begin{split}
    {f}&_{X,\Theta_{x}}( {x,\theta_{x}} ) = x  {f}_{X_{re},~X_{im}}\left( {x{\cos\theta_{x}},~x{\sin\theta_{x}}} \right)\\
    &=\frac{x}{4\pi^{2}}{\int_{0}^{\infty}{\omega~{\exp( - \gamma\omega)}~{\int_{0}^{2\pi}{{\exp\left\lbrack j\omega\left( {\delta_{1}{\cos\theta_{\omega}} + \delta_{2}{\sin\theta_{\omega}}} \right) \right\rbrack}}}~{\exp\left\lbrack j\omega x\left( {{\cos\theta_{x}}{\cos\theta_{\omega}} + {\sin\theta_{x}}{\sin\theta_{\omega}}} \right) \right\rbrack}~d\theta_{\omega}}}~d\omega
\end{split}
\label{eq.deviation}
\end{equation}
Using the identity $\cos{A}\cos{B}+\sin{A}\sin{B}=\cos{(A-B)}$ and integrating over $\mathrm{\Theta}_x$ we obtain the marginal pdf of signal amplitude X.
\begin{equation}
\begin{split}
    {f}&_{X}(x) = {\int\limits_{\theta_{x}}^{~}{{f}_{X,~\Theta_{x}}\left( {x,~\theta_{x}} \right)~d\theta_{x}}} \\
    &= \frac{x}{4\pi^{2}}{\int_{0}^{\infty}{{\mathit{\omega~}\exp}( - \gamma\omega)}}{\int_{0}^{2\pi}{{\exp\left\lbrack j\omega\left( {\delta_{1}{\cos\theta_{\omega}} + \delta_{2}{\sin\theta_{\omega}}} \right) \right\rbrack}{\int_{0}^{2\pi}{{\exp\left\lbrack j\omega x{\cos\left( \theta_{x} - \theta_{\omega} \right)} \right\rbrack}~d\theta_{x}}}~d\theta_{\omega}}}~d\omega
\end{split}
\end{equation}
By invoking the integral form of the zeroth-order Bessel function of the first kind $J_0(x)=\frac{1}{2\pi}\int_{0}^{2\pi}{\exp{\left(jx\sin{\theta}\right)}\ d\theta}$, the above pdf can be reduced into:
\begin{equation}
    {f}_{X}(x) = x{\int_{0}^{\infty}{{{\mathit{\omega~}\exp}( - \gamma\omega)}~{J_{0}(\omega\delta)}~J_{0}(\omega x)~d\omega}}~~~~~~~~~(x > 0)
    \label{eq.CRpdf}
\end{equation}
$\delta=\sqrt{\delta_1^2+\delta_2^2}$ becomes a unified location parameter. As one may have noticed, expression~(\ref{eq.CRpdf}) for Cauchy-Rician distribution derived here is very much similar to the generalized Rayleigh model in \cite{kuruoglu2004modeling}, which corresponds to an analytical parameter estimator using the second-kind statistics \cite{achim2006sar}. Unfortunately, introduction of the additional Bessel term $J_{0}(\omega\delta)$ within the the integral renders the original estimation method useless, and analytical solution to the Cauchy-Rician model remains an open problem.

It is worth noting that the original Cauchy-Rician pdf as proposed in \cite{karakucs2022cauchy} adopts both an intermediate and a final form that are distinct from expression~(\ref{eq.CRpdf}) proposed in this work, they are listed as follows.
\begin{equation}
    f_{X}(x) = \frac{x}{2\pi}{\int_{0}^{2\pi}{{\int_{0}^{\infty}{{{\mathit{\omega~}\exp}\left( {- \gamma\omega} \right)}~J_{0}\left( {\omega\sqrt{\left( {\delta_{1} + x{\cos\theta_{x}}} \right)^{2} + \left( {\delta_{2} + x{\sin\theta_{x}}} \right)^{2}}} \right)d\omega}}~d\theta_{x}}}
\label{eq.KarakusIntermediate}
\end{equation}
\begin{equation}
    f_{X}(x) = \frac{x\gamma}{2\pi}{\int_{0}^{2\pi}\frac{d\theta}{\left\lbrack \gamma^{2} + x^{2} + \delta^{2} - 2x\delta{\cos\left( \theta - \theta_{\delta} \right)} \right\rbrack^{3/2}}}
    \label{eq.KarakusFinal}
\end{equation}
$\theta_{\delta}$ is defined as ${\tan\theta_{\delta}} = \delta_{2}/\delta_{1}$. The intermediate expression~(\ref{eq.KarakusIntermediate}) which contains two integrals and one Bessel function is obtained by integrating out $\theta_\omega$ before moving on to resolve the amplitude distribution as suggested in Equation~(\ref{eq.deviation}). The equivalence between the it and Equation~(\ref{eq.CRpdf}) is readily proved using the following identity \cite{gradshteyn2014table}
\begin{equation}
    {\int_{0}^{\pi}{\left( {\sin x} \right)^{2\nu}\frac{J_{\nu}\left( \sqrt{\alpha^{2} + \beta^{2} + 2\alpha\beta{\cos x}} \right)}{\left( \sqrt{\alpha^{2} + \beta^{2} + 2\alpha\beta{\cos x}} \right)^{\nu}}~dx}}=2^{\nu}\sqrt{\pi}~\Gamma\left( {\nu + \frac{1}{2}} \right)\frac{J_{\nu}(\alpha)}{\alpha^{\nu}}\frac{J_{\nu}(\beta)}{\beta^{\nu}},~~~~~~~~~\left\lbrack {{\rm Re}~\nu > - \frac{1}{2}} \right\rbrack.
\end{equation}
Despite assuming a seemingly simplified form, the final form shown in Equation~(\ref{eq.KarakusFinal}) from reference \cite{karakucs2022cauchy} is basically an elliptic integral, whose analytical solution is far from easy to obtain. To prove this statement, assume a substitute variable $t = {\cos\theta}$, and its differential counterpart $d{\cos\theta} = - {\sin\theta}~d\theta$, Equation~(\ref{eq.KarakusFinal}) can then be transformed into
\begin{equation}
    f_{X}(x) = \frac{x\gamma}{\pi}\frac{1}{\left( {2x\delta} \right)^{3/2}}{\int_{- 1}^{1}{\frac{1}{\left( {\frac{\gamma^{2} + x^{2} + \delta^{2}}{2x\delta} - t} \right)^{3/2}}\frac{1}{\sqrt{1 - t^{2}}}~dt}}.
\end{equation}
Invoking the following identity \cite{gradshteyn2014table}
\begin{equation}
\begin{split}
    {\int_{u}^{b}\frac{dx}{\sqrt{\left( {a - x} \right)^{3}\left( {b - x} \right)\left( {x - c} \right)}}} = \frac{2}{\left( {a - b} \right)\sqrt{a - c}}E\left( {p,q} \right)\\
    p = {\arcsin\sqrt{\frac{(a - c)(b - u)}{(b - c)(a - u)}}},~q = \sqrt{\frac{b - c}{a - c}},~\left\lbrack {a > b > u \geq c} \right\rbrack
\end{split}
\end{equation}
Equation~(\ref{eq.KarakusFinal}) is reduced to 
\begin{equation}
    f_{X}(x) = \frac{2\gamma x~\boldsymbol{E}\left( {\sqrt{\frac{4r\delta}{\gamma^{2} + (r + \delta)^{2}}}} \right)}{\pi\left\lbrack \gamma^{2} + (r - \delta)^{2} \right\rbrack\sqrt{\gamma^{2} + (r + \delta)^{2}}}
    \label{eq.elliptic},
\end{equation}
where $E\left( {\psi,k} \right)$ is called "elliptic integral of the second kind," while $\boldsymbol{E}(k) = ~E\left( {\frac{\pi}{2},~k} \right)$ with bold font is "complete elliptic integral of the second kind". Both are well-established tabled functions.

It is certainly an optimal choice to use expression~(\ref{eq.KarakusFinal}) for MCMC-based parameter estimation since a rejection ratio involving the pdf has to be calculated for every data sample, but being an elliptic integral makes the expression promising when trying to develop an analytical parameter estimation method.

\subsection{Proposal of method of algebraic moments}
Conventional method of moments forms equations based on central moments (mathematical expectation of a power function). In the case of Cauchy-Rician, however, power function fails to provide enough analytical representations of moments other than for $E[x^{-1}]$. Other substitutes such as log moments or log-cumulants \cite{tison2004new} also fail to accomplish the job. This paper generalizes the method of moments by introducing statistical expectations of more complicated functions, such as $E\left[\left(x^2+a^2\right)^{-\frac{1}{2}}\right]$ and $E\left[\left(x^2+a^2\right)^{-\frac{3}{2}}\right]$. Despite their intricacy compared to the traditional power function, these algebraic functions performs surprisingly well as they produce moments with better simplicity, partially because of their resemblance to the Cauchy pdf. Derivation of these moments are presented as follows.

For the first algebraic moment, 
take $f\left(x\right)=\left(x^2+a^2\right)^{-\frac{1}{2}}$, where $a$ is an arbitrary number with certain restrictions (defined later by identities). The expectation of this estimator function according to the Cauchy-Rician model is
\begin{equation}
    E\left\lbrack \frac{1}{\sqrt{x^{2} + a^{2}}} \right\rbrack={\int_{0}^{\infty}{\omega~{\exp( - \gamma\omega)}~J_{0}\left( {\omega\delta} \right)~{\int_{- \infty}^{\infty}{\frac{x}{\sqrt{x^{2} + a^{2}}}J_{0}\left( {\omega x} \right)~dx}}~d\omega}}.
\end{equation}
Invoking the following identity \cite{gradshteyn2014table}, and noting that integrand x as signal amplitude is non-negative
\begin{equation}
    {\int_{0}^{\infty}{\frac{x}{\sqrt{x^{2} + a^{2}}}~J_{0}\left( {xy} \right)~dx}} = \frac{e^{- ay}}{y},~~~~~~~~~~\lbrack y > 0,{\rm Re}~a > 0\rbrack,
\end{equation}
we get rid of the integral on x:
\begin{equation}
    E\left\lbrack \frac{1}{\sqrt{x^{2} + a^{2}}} \right\rbrack = {\int_{0}^{\infty}{{\exp\left\lbrack - (\gamma + a)\omega \right\rbrack}~J_{0}\left( {\omega\delta} \right)~d\omega}}~~~~~~~~~~\lbrack {\rm Re}~a > 0\rbrack.
\end{equation}
Applying the following identity \cite{gradshteyn2014table},
\begin{equation}
    {\int_{0}^{\infty}{e^{- ax}~J_{\nu}\left( {bx} \right)~dx}} = \frac{b^{- \nu}\left\lbrack {\sqrt{a^{2} + b^{2}} - a} \right\rbrack^{\nu}}{\sqrt{a^{2} + b^{2}}}~~~~~~~~~~\left\lbrack {\rm Re}~\nu > - 1,{\rm Re}~\left( {a \pm ib} \right) > 0 \right\rbrack
\end{equation}
the expectation of the estimator function becomes
\begin{equation}
    E\left\lbrack \frac{1}{\sqrt{x^{2} + a^{2}}} \right\rbrack = \frac{1}{\sqrt{(\gamma + a)^{2} + \delta^{2}}},~~~~~~~~~~\lbrack {\rm Re}~a > 0\rbrack.
    \label{eq.moment_one}
\end{equation}

For the second algebraic moment, assume $f\left(x\right)=\left(x^2+a^2\right)^{-\frac{3}{2}}$, invoke identity \cite{gradshteyn2014table}
\begin{equation}
    {\int_{0}^{\infty}{\frac{x}{\left( {x^{2} + a^{2}} \right)^{3/2}}~J_{0}\left( {xy} \right)~dx}} = \frac{e^{- ay}}{a},~~~~~~~~~~\lbrack y > 0,{\rm Re}~a > 0\rbrack,
\end{equation}
to simplify the expectation 
\begin{equation}
    E\left\lbrack \left( {x^{2} + a^{2}} \right)^{- \frac{3}{2}} \right\rbrack =\frac{1}{a}{\int_{0}^{\infty}{\omega~{\exp\left\lbrack - (\gamma + a)\omega \right\rbrack}~J_{0}\left( {\omega\delta} \right)~d\omega}}~~~~~~~~~~\lbrack {\rm Re}~a > 0\rbrack
\end{equation}
Invoke the following identity \cite{gradshteyn2014table}, and please note that for the first order Legendre function we have $P_1\left(x\right)=x$.
\begin{equation}
    {\int_{0}^{\infty}{x^{\mu - 1}{\exp( - ax)}~J_{\nu}\left( {bx} \right)~dx}} =\left( {a^{2} + b^{2}} \right)^{- \frac{\mu}{2}}~\Gamma\left( {\nu + \mu} \right)P_{\mu - 1}^{- \nu}\left( \frac{a}{\sqrt{a^{2} + b^{2}}} \right)~~~~~~~~~~\left\lbrack a > 0,~b > 0,{\rm Re}~\left( {\nu + \mu} \right) > 0 \right\rbrack
\end{equation}
Hence, the final integral is removed:
\begin{equation}
    E\left\lbrack \left( {x^{2} + a^{2}} \right)^{- \frac{3}{2}} \right\rbrack = \frac{\gamma + a}{a\left\lbrack {(\gamma + a)^{2} + \delta^{2}} \right\rbrack^{3/2}}~~~~~\lbrack a > 0\rbrack.
    \label{eq.moment_two}
\end{equation}

Combination of the Equations~(\ref{eq.moment_one}) and (\ref{eq.moment_two}) automatically leads to a compact parameter estimator: Calculate the expectations of $E_1=E\left[\left(x^2+a^2\right)^{-\frac{1}{2}}\right]$ and $E_2=E\left[\left(x^2+a^2\right)^{-\frac{3}{2}}\right]$ from empirical data, and the estimated values of $\gamma$ and $\delta=\sqrt{\delta_1^2+\delta_2^2}$ can be solved for:
\begin{equation}
    \left\{ {\begin{matrix}
{\hat{\gamma} = a\left( {\frac{E_{2}}{E_{1}^{3}} - 1} \right)} \\
{\hat{\delta} = \sqrt{E_{1}^{- 2} - \left( \hat{\gamma} + a \right)^{2}}} \\
\end{matrix}~~~~~~~~~~\lbrack a > 0\rbrack} \right.
\end{equation}

A careful reader may find that Equation~(\ref{eq.moment_one}) or (\ref{eq.moment_two}) along with two different constant $a$ would also yield a functional estimator. But the two-moment estimator is reported slightly more stable than its single-moment counterpart, the equations in the former being less correlated.

\section{Experimental studies} 

\subsection{Experiments on synthetic data}
To prove the stability of the parameter estimator, the method is first applied to a 30 by 40 by $4\times10^{4}$ matrix of synthetic data, with the scale parameter $\gamma$ ranging from [5:5:150], and unified location parameter $\delta$ ranging from [5:5:200], and $4\times10^{4}$ data samples for each combination of parameters. This configuration of data is capable of simulating most types of surfaces in actual radar image classification, which is commonly carried out in small segmented image patches of a few hundred pixels wide.
The synthetic data are generated using methods suggested in \cite{ferguson1962representation}, and $\delta_1$ and $\delta_2$ for the real and imaginary parts of an synthetic complex signal are randomized for each $\delta$. According to repeated experiments, a value of the arbitrary constant $a$ close to the mean of the population yields the best performance.

\begin{figure*}[ht]
    \centering
    \begin{minipage}[h]{0.45\linewidth}
    \centerline{\includegraphics[width=8cm]{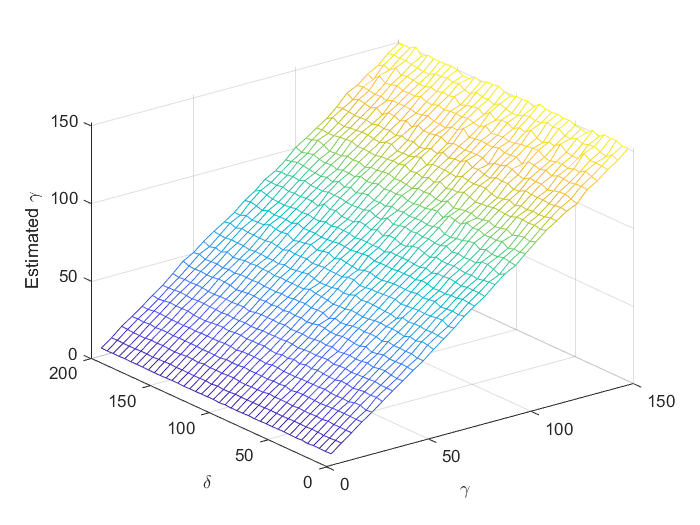}}
    \centerline{(a) estimated $\gamma$}\smallskip
    \end{minipage}
    \begin{minipage}[h]{0.45\linewidth}
    \centerline{\includegraphics[width=8cm]{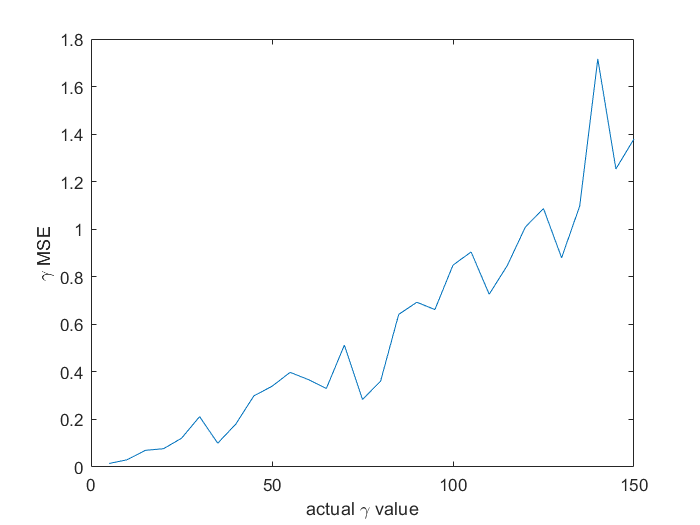}}
    \centerline{(b) $\gamma$ MSE}\smallskip
    \end{minipage}
    \begin{minipage}[h]{0.45\linewidth}
    \centerline{\includegraphics[width=8cm]{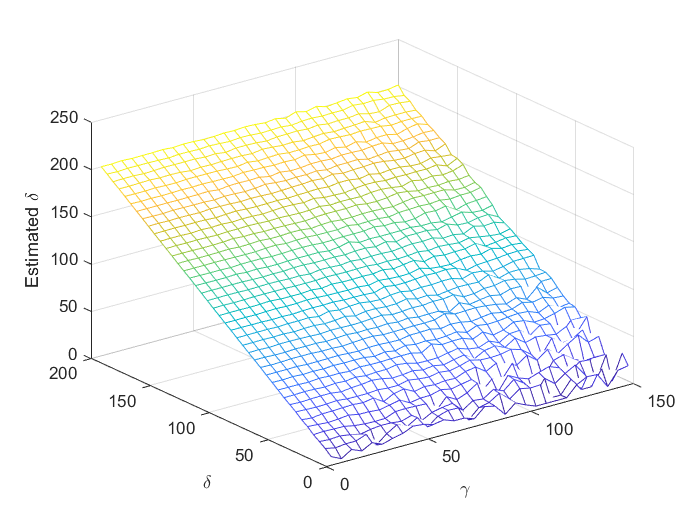}}
    \centerline{(c) estimated $\delta$}\smallskip
    \end{minipage}
    \begin{minipage}[h]{0.45\linewidth}
    \centerline{\includegraphics[width=8cm]{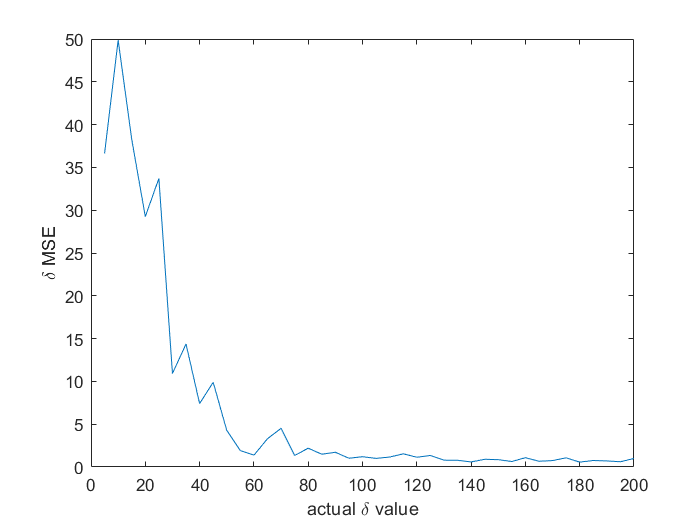}}
    \centerline{(d) $\delta$ MSE}\smallskip
    \end{minipage}
    \caption{Experimental results on synthetic data}
\end{figure*}  

Fig. 1 shows the estimated value of one parameter against variation of the other. A relatively low MSE ensures the estimator proposed in this work provides a faithful and robust estimation while covering a large range of parameter value. A significant advantage of this method comparing to  competing MCMC algorithm is the drastic reduction of computational power. When conducted with a laptop-based Intel i7-10750H CPU using MATLAB, the method of algebraic moment algorithm takes an average of 68 milliseconds to process a data set of $200 \times 200$ pixels, while the MCMC-based algorithm \cite{karakucs2022cauchy} takes more than an hour to finish, since the complicated pdf of the Cauchy-Rician (Equation~(\ref{eq.CRpdf})) is continuously calculated for every pixel.  

The reader is warned that the estimation of $\delta$ starts oscillate when $\delta$ is extremely small and $\gamma$ extremely large. Fortunately, empirical data generally are out of this unstable range. 

\subsection{Experiments on real empirical data}
\label{sec:typestyle}
\begin{figure}[ht]
    \centering
    \includegraphics[width=0.9\columnwidth]{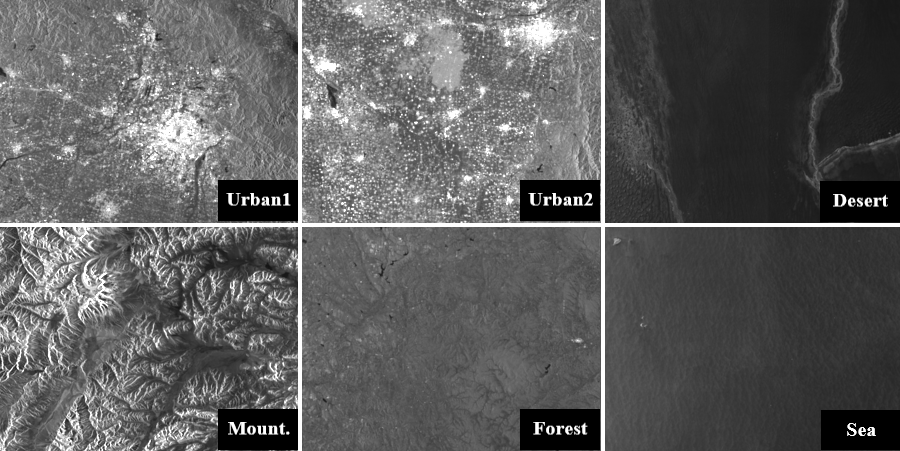}
    \caption{SAR images of various scenes used in experiments}
\end{figure}
The parameter estimation method is further applied to empirical data of C-band SAR images captured by Sentinel-1 satellite. Images of various scenes are taken into consideration, including that of urban area, desert, mountain, forest, and sea scenes, as shown in Fig. 2. Performance of the Cauchy-Rician model is evaluated using Kullback-Leibler Divergence (KL-div), and is compared with existing models including Rician, Weibull, log-normal, and ${\mathcal{G}_{0}}$ \cite{frery1997model}. Winner models for each scene are highlighted in bold font.

\begin{table}
    \centering
    \begin{tabular}{c|c c c c c}
    \hline
    KL-div & CR & Rician & Weibull & log-n & $\mathcal{G}_{0}$ \\
    \hline
    Urban1 & \textbf{0.1498} & 0.3066 & 0.2069 & 0.2589 & 0.2562 \\
    Urban2 & \textbf{0.1825} & 0.2949 & 0.2053 & 0.2560 & 0.2501 \\
    Urban3 & \textbf{0.1897} & 0.2097 & 0.1987 & 0.2267 & 0.2037 \\
    Urban4 & \textbf{0.1352} & 0.2088 & 0.1641 & 0.2153 & 0.2212 \\
    Urban5 & \textbf{0.1200} & 0.2875 & 0.1976 & 0.2474 & 0.2435 \\
    Urban6 & \textbf{0.1169} & 0.2148 & 0.1832 & 0.2221 & 0.1908 \\
    Desert & \textbf{0.1784} & 0.1938 & 0.1817 & 0.2321 & 0.2317 \\
    Mount. & \textbf{0.0998} & 0.3653 & 0.2365 & 0.3062 & 0.3193 \\
    Forest & 0.3236 & 0.2022 & 0.1380 & 0.1932 & \textbf{0.1321} \\
    Sea & 0.3659 & 0.1637 & \textbf{0.1256} & 0.1747 & 0.1337 \\
    \hline
    \end{tabular}
    \caption{KL-div measurement of modelling performance, CR: Cauchy-Rician, log-n: log-normal}
    \vspace{-0.3cm}
    \label{tab:my_label}
\end{table}
Table 1 shows that the Cauchy-Rician model performs especially well in modelling heterogeneous data of urban and mountainous areas, but fails to excel for more homogeneous data populations such as sea and forest scenarios.
\begin{figure}[ht]
    \centering
    \includegraphics[width=0.9\columnwidth]{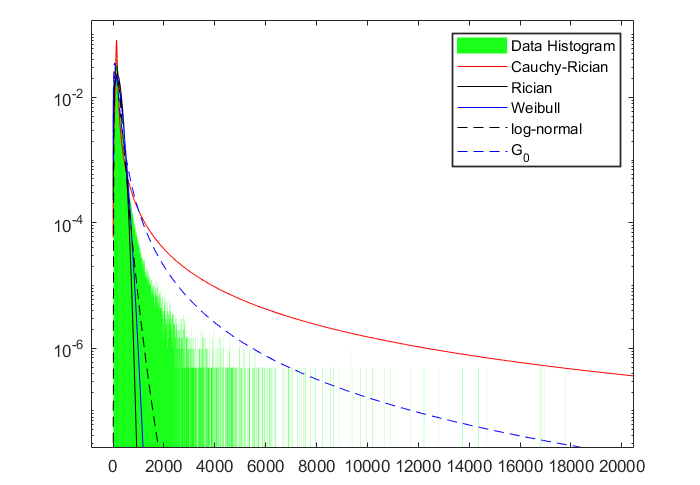}
    \caption{Goodness of fit of various models on Urban1 scene }
\end{figure}

Fig. 3 shows the goodness of fit of SAR image "Urban1", where normalized histogram is compared against various models including Cauchy-Rician, Rician, Weibull, log-normal, and $\mathcal{G}_{0}$. The Cauchy-Rician model produces a better description than other models of the thick tail of heterogeneous urban scenes (when x-axis exceeds 14000), in parallel to the results in Kullback-Leibler Divergence test. However, it is also apparent that Cauchy-Rician's superior representation of the tail happens at the cost of a over-shoot in the peak ($x\approx500$), this suggest that Cauchy-Rician is a bit too impulsive even for extremely heterogeneous urban scene image.

\section{Conclusions}
\label{sec:foot}

This paper has proposed and verified a compact analytical parameter estimator for the Cauchy-Rician model based on a generalized method of (algebraic) moments that significantly reduced the computation time compared to preceding works. Performance of the estimation method together with the Cauchy-Rician model have been demonstrated by experiments on both synthetic data and extremely heterogeneous SAR images. In particular, the Cauchy-Rician model performs exceptionally well in modelling SAR images of urban and mountainous areas, primarily due to the model's inherent heavy-tailed character arising from the  Cauchy distribution which characterizes impulsive data populations successfully. 
 Our future work will explore other applications of the Cauchy-Rician model 
 and its modification for asymmetric data distributions \cite{kuruoglu03}.

\bibliography{mybibfile}

\end{document}